\documentclass[journal=nalefd,manuscript=article]{achemso}

\usepackage{graphicx}
\graphicspath{{Figures/PNG files}{./}}
\usepackage{dcolumn}
\usepackage{bm}
\usepackage{float}
\usepackage{hyphenat}
\usepackage{natbib}
\usepackage{hyperref}
\usepackage[dvipsnames]{xcolor}
\usepackage{xcolor}
\usepackage{amsmath}

\usepackage{tikz}

\usepackage{comment}

\newcommand{\brownmark}[1] {\color{purple}\textbf{#1}\color{black}\normalsize}

\newcommand{\inl}{QuantaLab, International Iberian Nanotechnology Laboratory (INL), Avenida Mestre Jos\'{e} Veiga, 4715-330 Braga, Portugal}

\newcommand{\san}{Istituto di Struttura della Materia, Consiglio Nazionale delle Ricerche (CNR-ISM), Division of Ultrafast Processes in Materials (FLASHit), Via Salaria Km 29.5, CP 10, I-00016 Monterotondo Stazione, Italy}

\title{Magneto-optical response of chromium trihalide monolayers: chemical trends}

\author{Alejandro Molina-S\'{a}nchez}
\affiliation{\inl}
\email{alejandro.molina@uv.es}
\author{Gon\c{c}alo 
Catarina}
\affiliation{\inl}
\author{Davide Sangalli}
\affiliation{\san}
\author{Joaqu\'{i}n 
Fern\'{a}ndez-Rossier
\brownmark{\footnote{On permanent leave from Departamento de Fisica Aplicada, Universidad de Alicante, San Vicente del Raspeig, 03690 Spain }}
}
\affiliation{\inl}

\date{\today}

\begin{document}

\begin{abstract}
Chromium trihalides (CrI$_3$, CrBr$_3$ and CrCl$_3$) form a prominent family of isostructural insulating layered materials in which 
ferromagnetic order has been observed down to the monolayer. Here we provide a comprehensive computational study of magneto-optical properties that are used as probes for the monolayer ferromagnetic order: magnetic circular dichroism and magneto-optic Kerr effect. Using a combination of density functional and  Bethe-Salpeter theories, we calculate both  the optical
absorption and the magneto-optical Kerr angle spectra, including both excitonic
effects and  spinorial wave functions. We compare the magneto-optical response
of the chromium trihalides series and we find that its strength is governed by 
the spin-orbit coupling of the ligand atoms (I, Br, Cl).
\end{abstract}

\maketitle

\section{Introduction}

The family of chromium trihalides plays a very prominent role in the research area of magnetic 2D crystals.
Ferromagnetic order down to the monolayer in a stand-alone 2D crystal was
first reported for a CrI$_3$  sample, back in 2017\cite{Huang2017,Klein2018}.
This discovery triggered an intensive research 
that has led to the discovery of ferromagnetic order in monolayers and few
layers of several compounds \cite{Gong2019,Gibertini2019}, including
CrBr$_3$\cite{Zhang2019a,Kim2019}, CrCl$_3$\cite{Cai2019} and many
others\cite{Gong2017,Fei2018,Deng2019}. The fabrication of Van der Waals
heterostructures integrating these newly discovered 2D ferromagnets with other
2D crystals\cite{Burch2018} has further fueled this research area and has motivated
several theory proposals of new spintronic devices\cite{Cardoso2018,Zollner2019} and topological phases \cite{Li2018}. 

Magneto-optical probes, such as Kerr effect and magnetic circular dichroism
(MCD), are  widely used to probe  the existence  of ferromagnetic order in
monolayers\cite{Huang2017,Sivadas2018a,Huang2018,Mak2019,Thiel2019,Sun2019}, as
well as to probe the valley polarization\cite{Mak2018} and magnetic proximity 
effects\cite{Seyler2018b}.  The detection limit of conventional SQUID 
magnetometry, down to 10$^{12}\, \mu_B$ \cite{Buchner2018},  sets a lower limit for the 
area of the monolayers in the range of $(100 \mu m)^2$,  whereas the 
typical flakes have linear dimensions smaller than $10\mu m$.   

Both the Kerr angle and MCD can be related to the transverse (Hall) ac
conductivity of the compounds, $\sigma_{xy}(\omega)$
\cite{Argyres1955,Guo1995,Gudelli2019}. At the microscopic level, this quantity
arises from Lorentz-type forces acting on the electrons. In the absence of an
external magnetic field,  $\sigma_{xy}$ is only non-zero when both spin-orbit interaction and the breaking of time reversal symmetry, inherent in the ferromagnetic order, are present \cite{Catarina2019a}. Spin-orbit coupling is also essential in magnetic 2D crystals as it brings magnetic anisotropy, that ensures the existence of magnetic order in two dimensions. In the case of CrI$_3$ it has been demonstrated that it is the spin-orbit coupling of the ligand, the iodine atom, the one that controls magnetic anisotropy\cite{Lado2017}.

So far, theoretical efforts have been focused on the excitonic effects on Kerr
angle in monolayer CrI$_3$\cite{Wu2019a}, or few-layers and bulk\cite{Gudelli2019}. Others 
\textit{ab initio} studies have described magnetic anisotropy and critical
temperatures of ferromagnetic 2D materials\cite{Torelli2018,Torelli2019},
localized surface waves in CrI$_3$ structures\cite{Pervishko2019}, and
structural properties\cite{Woo2019}. Nevertheless, a theoretical study focused on the chemical trends of the magneto-optical
properties of the family of chromium trihalides, including excitonic effects,
is still missing.

With this background, it is natural to enquire which atom, either Cr or the ligand, provides the spin-orbit interaction that controls the magnet in the CrX$_3$ family, with X$=$I,Br,Cl.  Therefore,  here we undertake an \textit{ab initio} study of the chromium trihalide family.  Our methodology includes a density functional theory (DFT) calculation to obtain ground state properties, extended with a GW and Bethe Salpeter calculation to compute the optical response, including the interplay between magnetism and spin-orbit coupling.  These methods, that permit to fully include the  excitonic effects, known to be very strong in 2D crystals,  have been employed to study the CrI$_3$ monolayers in a recent publication \cite{Wu2019a}. 
Our work permits to carry out a comparative analysis of the optical and magneto-optical response of the CrX$_3$ series.

\begin{figure}[t]
\begin{center}
\includegraphics[scale=0.40]{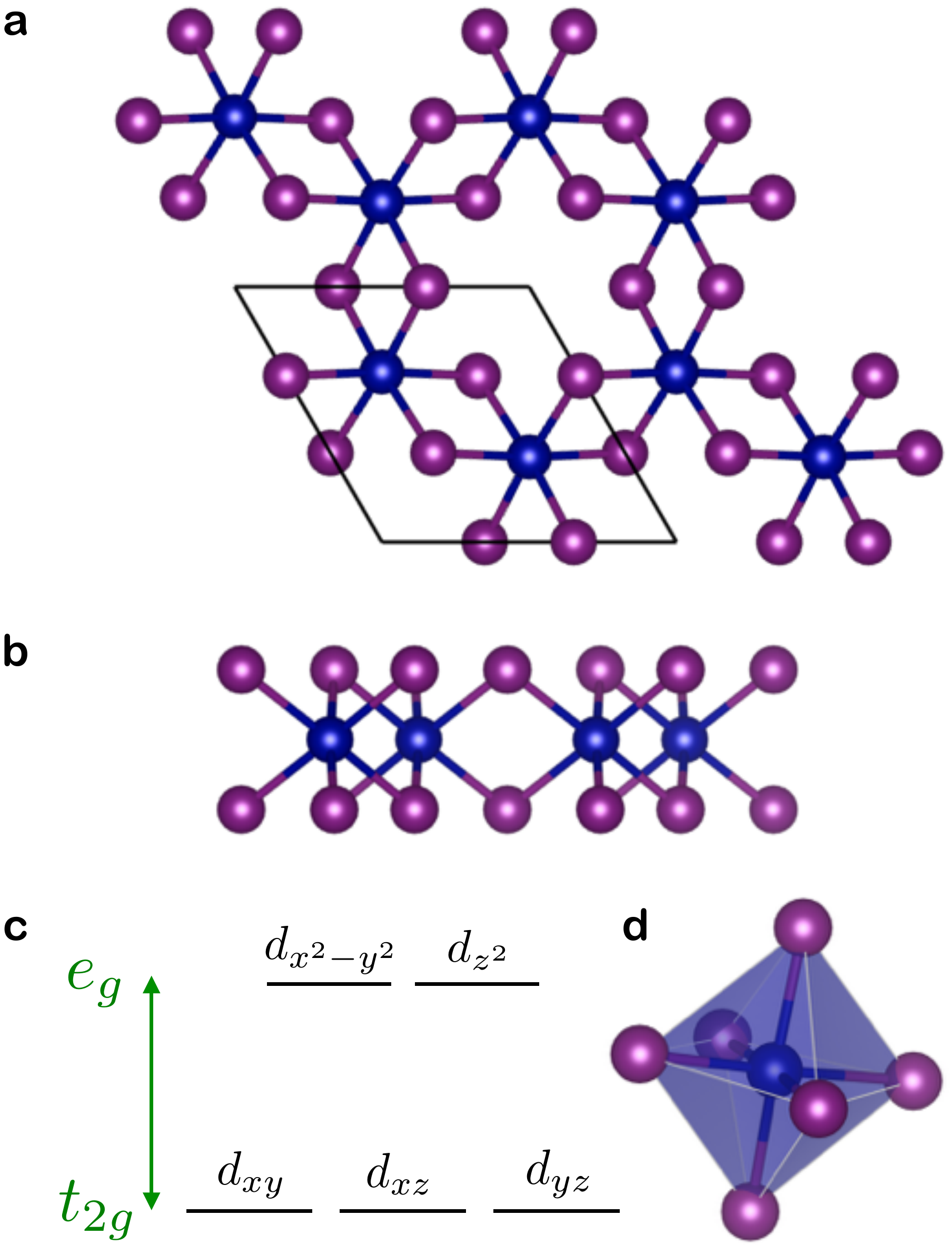}
\end{center}
\caption{Crystalline structure of monolayer chromium trihalides (point symmetry $D_{3d}$). (a-b) Top and lateral view. (c-d) Scheme of the energy alignment of Cr $d$-orbitals as a result of the crystal-field splitting produced by the
distortion of the octahedral environment.}
\label{structure}
\end{figure}

\section{Methods\label{methods}}

\subsection{Ab initio and many-body perturbation theory}

In this section we briefly describe the \textit{ab initio} methodology used to compute the ground state properties of the CrX$_3$ and their optical response. 
The \textit{ab initio} calculations of the electronic structure of the monolayer CrI$_3$, CrBr$_3$ and CrCl$_3$ have been performed using Quantum Espresso \cite{Giannozzi2009}. We have employed the local density approximation (LDA) plus on-site Hubbard correction with
values $U=1.5$ eV and Hund's exchange interaction $J=0.5$ eV\cite{Baidya2018}.
We have included spin-orbit interaction with spinorial wave functions, using
norm-conserving fully relativistic pseudopotentials. The pseudopotentials of Cr atom include 
semi-core valence electrons and have been generated with \texttt{ONCVPSP} \cite{Hamann2013} 
and \texttt{PSEUDODOJO} \cite{VanSetten2018}. The electronic density converges with an energy 
cutoff of 87 Ry and a $\bf{k}$-grid of $12\times 12\times 1$. We use a slab model with a 
20 \AA\space vacuum thickness to avoid interactions between periodic images.
Since LDA underestimated the bandgap, we compute the GW band-gap corrections on top of the LDA+U.
Let us label (GW+U)\@(LDA+U) the final band-gap, since keep the U correction for the correlated 
$d$ electrons, which is not accounted for by the GW self--energy, on top of GW.
Other choices are of course possible, like for example the update of the U correction 
in the GW framework, or the removal of the U correction to the simpler scheme GW\@(LDA+U).
A discussion on the differences between these approaches is however beyond the goal of the present manuscript.

In order to obtain a realistic optical response, we have included excitonic effects, fundamental in semiconducting 
2D materials \cite{Olsen2016}. The excitonic effects are taken into accout via the Bethe-Salpeter Equation (BSE) as implemented in Yambo\cite{Marini2009,Sangalli2019}.
The excitonic spectra and the dielectric function has been converged with a 
$\bf{k}$-grid of $15\times 15\times 1$ and a dielectric cut-off of 5 Ry. The dynamical
screening effect has been included using the plasmon-pole approximation \cite{Larson2013}.
Since GW corrections are computed only at the band--gap, we used a scissor operator to extrapolate 
them to the whole $15\times 15\times 1$ $\bf{k}$-grid in the BSE calculations.
This choice does not alter the final result (we have checked that the GW correction is a rigid-shift of the LDA electronic structure) 
but reduces considerably the computational effort. The Coulomb interaction is truncated by using 
the Coulomb cut-off technique\cite{Rozzi2006}.
Finally we use the ``Covariant approach''~\cite{Sangalli2019} for the dipole matrix elements to account 
for the non local $U$ term included in the DFT  Hamiltonian. This is crucial to
obtain correct intensities in the absorption and thus a correct estimation of the Kerr angles.

\subsection{MCD and Kerr angle in terms of the dielectric tensor}

An effective dielectric tensor is defined as
\begin{equation}
\varepsilon=1+\frac{4\pi\alpha_{2D}}{d},
\end{equation}
where $\alpha_{2D}$ is
the polarizability per surface unit and $d$ the 2D-material thickness.
We have assumed a thickness $d_{\text{CrX}_3}=0.66$ nm for all the monolayers.
It is calculated with the Yambo code, including excitonic effects and local-field effects\cite{Sangalli2019}. From the 
effective dielectric tensor we can define the absorbance for linearly ($A_x$) and circularly ($A_{\pm}$) polarized light as\cite{Bernardi2013,Thygesen2017}:

\begin{equation}
A_x = \Im(\varepsilon_{xx})
\frac{\omega d}{c},\quad A_{\pm} = \Im(\varepsilon_{xx}\pm i\varepsilon_{xy})\frac{\omega d}{c} ,
\end{equation}
which is independent from $d$ and thus a property of the 2D material.

From the effective dielectric tensor we can also obtain the Kerr angle, defined
as the change to light polarization reflected from a magnetized surface. The expression for the Kerr angle is derived using the standard Fresnel formalism, assuming normal incidence of linearly polarized light (polar geometry).
In order to account for the effect of a substrate, we consider a stratified medium where the magnetic material, with thickness $d_{\text{CrX}_3}$, is placed between air and a semi-infinite dielectric substrate.
In the case of SiO$_2$ substrate, the relative permittivity at the relevant frequencies is $\varepsilon_r=2.4$~\cite{Constant2016}. 
The effective dielectric tensor of $CrX_3$ is diagonal due to the symmetries of the lattice.
In presence of a magnetization along the $z$-axis and spin--orbit coupling, it assumes the form

\begin{equation}
\varepsilon = 
\begin{pmatrix}
\varepsilon_{xx} & \varepsilon_{xy} & 0 \\
-\varepsilon_{xy} & \varepsilon_{xx} & 0 \\
0 & 0 & \varepsilon_{zz}
\end{pmatrix}.
\label{tensor}
\end{equation}
Taking advantage of this anti--symmetric form, the Kerr angle can be derived as\cite{Catarina2019a}:

\begin{equation}
    \theta_K = -\frac{1}{2} \arg 
    \left(\frac{r_+}{r_-} \right),
   \label{kerr}
\end{equation}
where $r_\pm$ are the reflection coefficients given by
\begin{eqnarray}
    r_{\pm}=\frac{1-n_{\mp}h(n_{\mp})}{1+n_{\mp}h(n_{\mp})},
    \label{r}
\end{eqnarray}
in which $n_\pm = \sqrt{\varepsilon_{xx} \pm i \varepsilon_{xy}}$ are the refractive indexes in the circular basis and
\begin{eqnarray}
    h(n_{\pm})&=&\frac{f(n_{\pm})-g(n_{\pm})}{f(n_{\pm})+g(n_{\pm})}, \\
    f(n_{\pm})&=&(n_{\pm}+\sqrt{\varepsilon_r})e^{-i\frac{\omega}{c}n_{\pm}d}, \\
    g(n_{\pm})&=&(n_{\pm}-\sqrt{\varepsilon_r})e^{i\frac{\omega}{c}n_{\pm}d}.
    \label{hfg}
\end{eqnarray}

Thus, the combination of equations (3-7) permits to 
relate the first-principles calculations of the dielectric tensor with the Kerr
angle, including the effect of a semi-infinite substrate.

\section{Ground state properties of the chromium trihalides \label{review}}

\subsection{Structural properties}

\begin{figure*}[t]
\begin{center}
\includegraphics[width=1.0\linewidth]{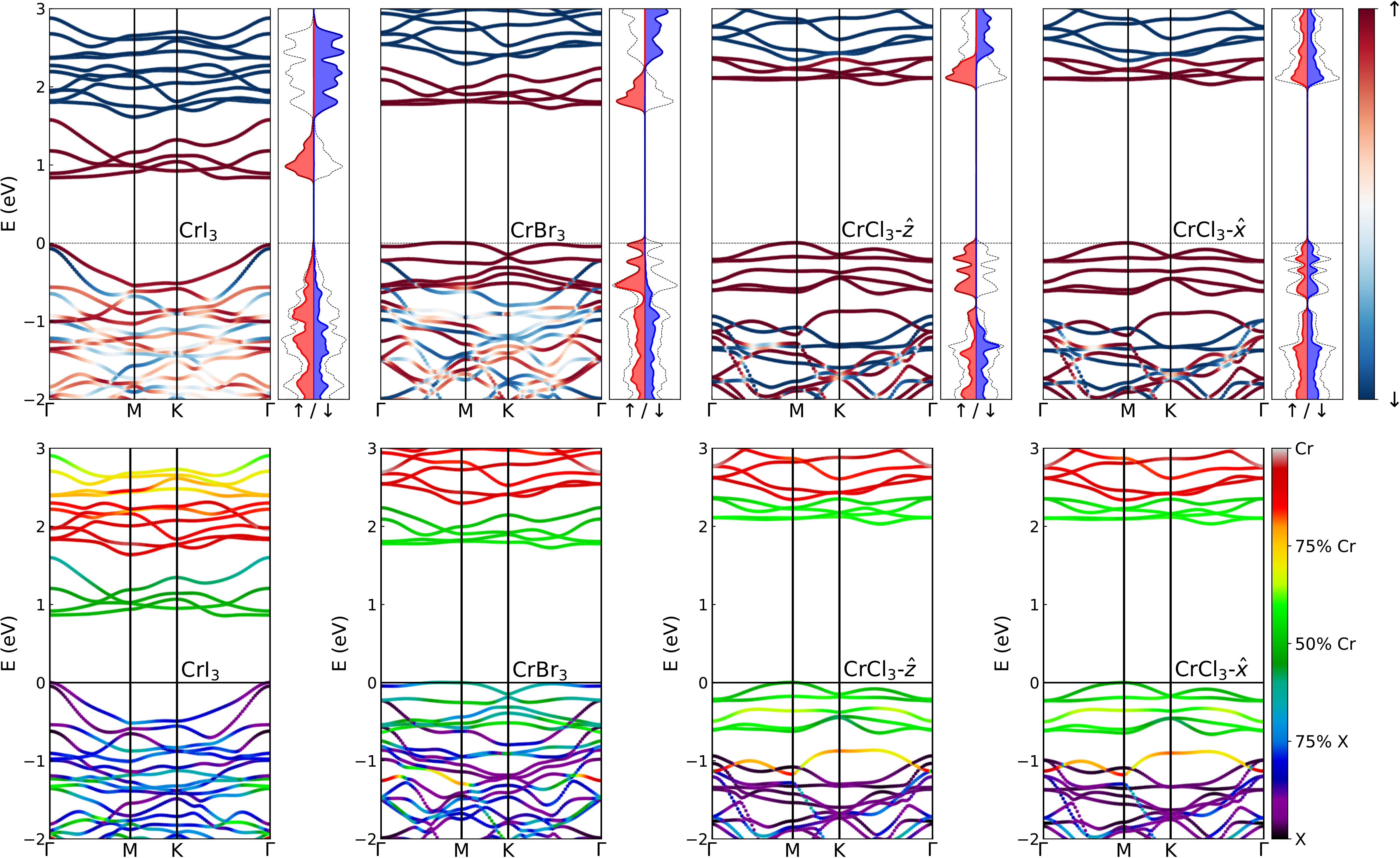}
\end{center}
\caption{Upper panel. Spin-projected band structure and density of states of
    chromium trihalides. Red is 100 \% spin-up and blue 100 \% spin-down.
Magnetization points in the positive $z$ direction for the first
three systems, in which case both the bands and the DOS have $S_z$ spin
projection. In the case of CrCl$_3$ with $\mathrm{\bm{M}}$ along $x$ direction, the bands 
are projected along $S_x$ whereas the density of
states is projected along $S_z$. Lower panel. Atom-projected band structures of idem materials.}
\label{bands}
\end{figure*}

The compounds of the Cr$X_3$ family share many structural, electronic and magnetic properties. They 
have identical crystalline structure (see Fig. \ref{structure}), a hexagonal lattice
with point group symmetry $D_{3d}$ \cite{McGuire2015}. The Cr atoms form a honeycomb 
lattice and are surrounded by the ligand's octahedron, as shown
in Fig. \ref{structure}d. As a result of the edge sharing geometry, first
neighbor Cr atoms share a pair of ligands, providing 90$^{\circ}$ pathways for 
super-exchange. Moreover, the three CrX$_3$  compounds have ferromagnetic order
down to the monolayer, with Curie temperatures of $T_{I}=45K$\cite{Huang2018}, $T_{Br}=34 K$\cite{Zhang2019a} 
and $T_{Cl}=17 K$\cite{Cai2019}. The magnetization easy axis is off-plane 
for CrI$_3$ and CrBr$_3$ and in plane for CrCl$_3$. Although 
not important in the case of monolayers, we also note that interlayer 
interactions in bulk are ferromagnetic for $X=I,Br$ and antiferromagnetic for $X=Cl$\cite{McGuire2017}. DFT calculations predict that, in the three compounds, the magnetic moment is hosted  predominantly in the Cr ions, that have $S=3/2$\cite{Streltsov2017}.

\subsection{Electronic properties: a simple model}

A qualitative understanding of the most salient electronic properties of the
CrX$_3$ family is derived from the atomic model for $d$ levels of Cr ions in the
octahedral environment of the ligands. In a fully ionic picture, the nominal
oxidation state of the Cr atoms is Cr$^{+3}$ and the ligand atoms are thereby in
the $X^{-1}$. The resulting crystal field originated by the charged Halide atoms
splits the single particle $d$ levels in an orbital triplet $t_{2g}$ and a
higher energy doublet $e_g$. Thus, the outermost electronic $d$ levels of the
Cr$^{3+}$ ions are occupied by 3 electrons, that occupy the single spin $t_{2g}$
triplet in order to minimize both orbital energy and Coulomb repulsion
(intra-atomic Hund's rule). We refer to the spin channel of the  occupied
$t_{2g}$ levels as {\em majority spin channel}. The energy arrangement of the
$d$ levels of the ionic model is shown in the scheme of Fig. \ref{structure}c,
and are confirmed by DFT results,\cite{Soriano2019} although the $t_{2g}$ levels
are strongly hybridized with the $p$ bands of the ligands.

The naive ionic model predicts that the outermost shell of the $X$ ions is a
full $p$ shell. Our DFT calculations confirm this picture. The resulting energy
bands that arise from the ligand $p$ shell and the spin majority Cr-$t_{2g}$
coexist in energy, are strongly hybridized, and constitute the valence band of
the CrX$_3$ family. The degree of hybridization will depend on the ligand
species. On the other hand, conduction bands are spin {\em majority} $e_g$
levels, hybridized with X $p$ orbitals. The unit cell of the CrX$_3$ has the formula Cr$_2$X$_6$. Therefore, there are 4 conduction bands, which are fully spin polarized. The lowest energy optical response is thus governed by transitions between the Cr-$t_{2g}$-X-$p$ valence band and the Cr-$e_g$ spin majority bands.

\begin{table*}[t]
    \centering
    \begin{tabular}{cccccc}
    \hline
    \hline
    Material & $E_{\text{LDA+U}}$ (eV) & $E_{\text{GW}}$ (eV) & $S_z$/$|S_z|$ & $E_B^{d}$ (eV) & $E_B^{b}$ (eV) \\
    \hline
    CrI$_3$             &  0.86 & 2.76 & 6.03/7.14 & 1.06 & 0.84 \\
    CrBr$_3$            &  1.75 & 4.45 & 6.00/6.68 & 2.05 & 1.99 \\
    CrCl$_3$-$\hat{z}$  &  2.04 & 5.47 & 6.00/6.44 & 2.62 & 2.57 \\
    CrCl$_3$-$\hat{x}$  &  2.09 & 5.47 & 6.00/6.44 & 2.63 & 2.57 \\
    \hline
    \hline
    \end{tabular}
    \caption{Summary of electronic and optical properties of CrX$_3$. Bandgaps as calculated within the LDA+U and GW method. Magnetization and excitonic binding energy
    of first dark ($E_B^{d}$) and bright ($E_B^{b}$) excitonic state.}
    \label{gaps}
\end{table*}

\vspace{1cm}

\subsection{Electronic properties: DFT results\label{DFT}}

The band structures and density of states (DOS) of CrX$_3$ are presented in Fig.
\ref{bands}. All the calculations are done with non collinear spin-orbit
coupling. The magnetization is oriented along the $z$-axis for CrI$_3$ and CrBr$_3$. In 
the case of CrCl$_3$ we consider magnetization along the $x$ axis, given that the easy-axis of this
compound is in-plane \cite{Wang2011,Cai2019}, and also address the case where a magnetization
along the $z$ axis is imposed. For the three materials the conduction band is
formed by $e_g$ spin majority bands, as expected from the ionic model of Fig.
\ref{structure}c. In the case of CrCl$_3$ and, to a lesser extent,  CrBr$_3$
the valence band is dominated by the $t_{2g}$ bands so that the energy gap is
inherently related to the crystal field splitting of the $t_{2g}-e_g$ manifold
in the spin majority channel. In the case of  CrI$_3$ the top of the valence
band becomes dominated by the $p$ band of Iodine. Therefore, the band gap in
CrI$_3$ is a metric of  the inter-atomic charge transfer energy overhead. The
bandgaps, as obtained with LDA+U, are  summarized in Table \ref{gaps}. It is
apparent that the bandgap correlates inversely with the nuclear charge of the
ligand, resulting in a larger bandgap for lighter ligands, as evidenced in Fig.
\ref{bands}. As expected, the crystal field splitting follows the sequence
$\Delta_{Cl} > \Delta_{Br} > \Delta_{I}$ on account of the larger
electronegativity of lighter ligands. On the other hand, the modulus of magnetization is mainly dominated by the chromium atoms\cite{Abramchuk2018} and it has similar values for the three compounds, as shown in Table \ref{gaps}.

The spin polarization of the valence band varies from material to material, on
account of the different degree of hybridization between the Cr $t_{2g}$ spin
majority levels and the $p$ orbitals of the ligands. Thus, the spin polarization
is found to be
nearly 100 \% spin-up for CrBr$_3$ and CrCl$_3$ \cite{Zhang2015}. The
stronger hybridization of Cr-I is also reflected in the absolute value of the
total magnetic moment of the unit cell, shown in Table \ref{gaps}, 
indicating that each iodine ion hosts a local moment, but with
antiparallel alignment among them, leading to a null iodine moment, when summed
up over the unit cell.\cite{Jiang2018} In the case of CrCl$_3$, the
difference between in-plane or out-of-plane magnetization barely modifies the band dispersion. 

In contrast, for the conduction bands, the $e_g$ bands are nearly 100 \% spin-up for all the compounds and with an increasing contribution from 
chromium $d$-orbitals when changing from iodine to chlorine. The implications on the optical properties of the valence band character will be discussed in the next Section.
The fact that the {\em spin majority} $e_g$ levels are below all the spin minority levels evidences
that the crystal field splitting is smaller than the intra-atomic exchange.

\section{Optical Properties: Dichroism and Kerr spectroscopy \label{MCD}}

From the band structures of the chromium trihalides series we can infer 
a strong dependence of the optical response on the ligand. Whereas 
the chromium atom hosts the magnetic moment, the ligand determines 
the strength of the spin-orbit interaction, the Cr-X hybridization 
of the valence band states and the magnetic anisotropy. Therefore, measuring 
absorbance (A) or photoluminescence (PL) 
of circularly polarized light,
as well as Kerr angle, permits to characterize
the magnitude of the magnetic anisotropy.

\begin{figure*}[t]
\begin{center}
\includegraphics[width=1.0\linewidth]{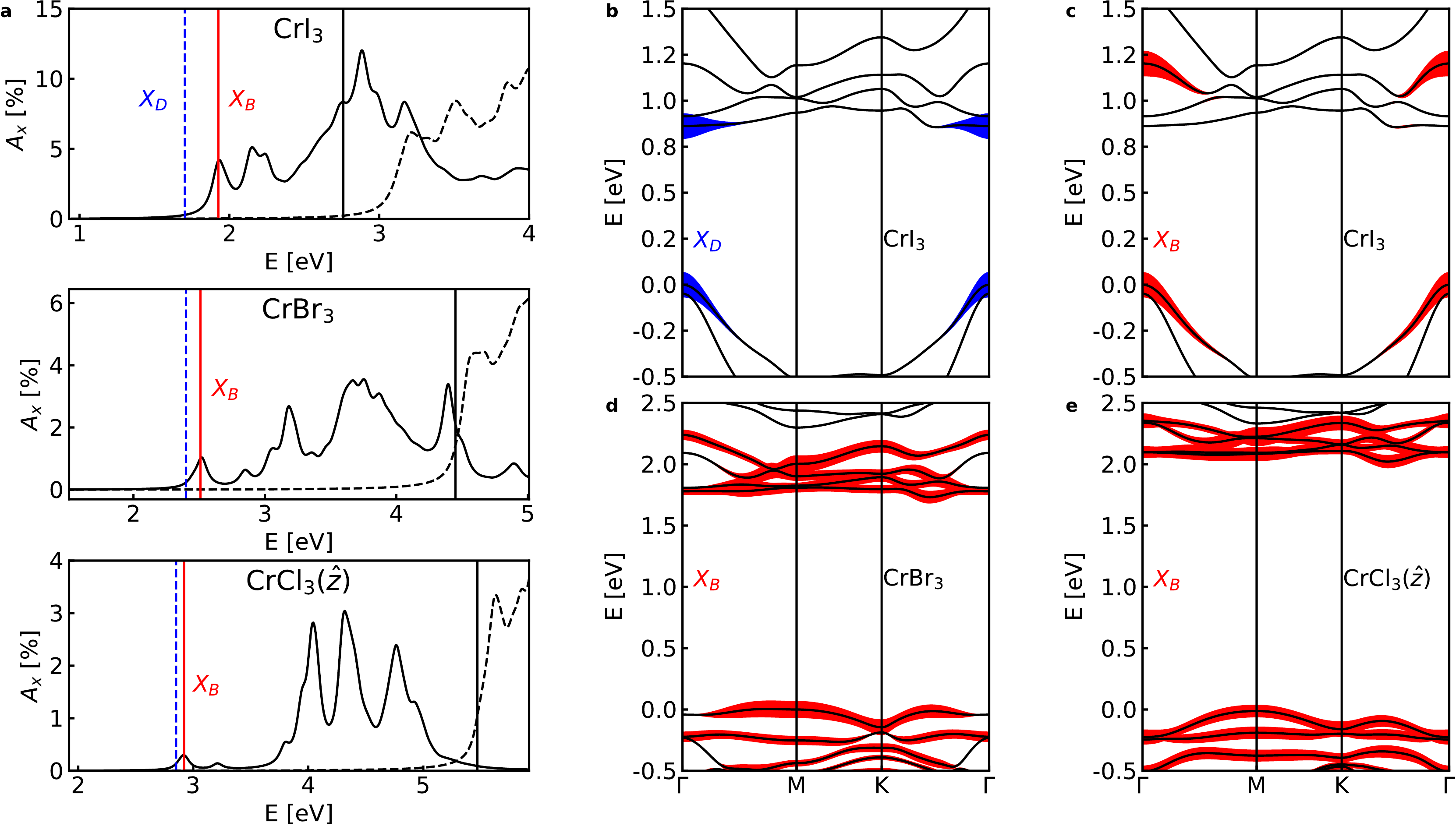}
\end{center}
\caption{(a) Absorbance of linearly polarized light of chromium trihalides with
magnetization out-of-plane. Solid (dashed) black lines correspond to calculations with (without) excitonic
effects. Vertical lines mark the energy of first bright (red
solid line) and dark (blue dashed line) exciton states together with the
electronic bandgap obtained with the GW method (black solid line). (b-e) Representation of the coefficients of the wave function of relevant excitonic states: (b-c) Dark and bright excitons ($X_D$ and $X_B$) of CrI$_3$; (d-e) Bright excitons of CrBr$_3$ and CrCl$_3$, respectively.}
\label{absorbance}
\end{figure*}

\subsection{Excitonic effects}
First, we expect strong excitonic effects on the optical properties in these 2D
materials. Figure \ref{absorbance} shows
the absorbance spectra of chromium trihalides with magnetization out-of-plane.
The difference in the absorption threshold energies calculated with (solid line)
and without (dashed line) excitonic effect, defines the exciton binding energy.
In the figure we mark with vertical lines both the optically dark (blue dashed
line) and the optically bright exciton binding energy.  Values for the  the
bright exciton binding energy are shown in Table \ref{gaps}. The exciton binding
energies are much higher than those of transition metal
dichalcogenides.\cite{Molina-Sanchez2013,Molina-Sanchez2015,Haastrup2018}. We
attribute this enhancement to the reduced dielectric screening
environment (see supporting information)\cite{Qiu2013,Latini2015,Noori2019,Raja2019}. 

The exciton binding energies are significantly higher for CrCl$_3$ and CrBr$_3$ than for CrI$_3$.  In order to  understand this difference, we  represent the electronic transitions that
contribute to the main excitonic peaks, by plotting the coefficients
of the exciton wave functions projected over the DFT bands in Figs.
\ref{absorbance}b-e (see refs. \cite{Molina-Sanchez2015, Paleari2018,Paleari2019} for details). In the case of
CrI$_3$, the excitonic wave function exhibits localized excitonic states in $\bf
k$-space, which implies extended wave function in real space. In contrast, for
CrBr$_3$ and CrCl$_3$, exciton wave functions expand over the full Brillouin
zone, implying an atomically localized exciton wave function. A larger
localization in momentum space implies a shorter electron-hole distance in real
space, and thereby, a larger exciton binding energy.  The ultimate origin of
this chemical trend is probably the stronger Cr-ligand hybridization in the
valence band of CrI$_3$. We note that our calculations set the absorbance 
threshold at 1.9 eV, whereas experiments found it at 1.5 eV \cite{Huang2017}.
The origin of the discrepancy is probably due to the role played by
electron-lattice coupling and a possible polaronic distorsion of the lattice.
The very large Stokes shift observed experimentally shows that electron.lattice
coupling is probably  important

The height of the absorption threshold at the bright exciton peak has also a very marked dependence on the ligand: absorption is strongest for CrI$_3$ and weakest for CrCl$_3$.  Given that the excitonic effect follows an inverse trend, we attribute the enhanced absorption of the CrI$_3$ to the large content of ligand wave function in the valence band states.

Moreover, we find as a common feature the existence of a dark ground excitonic
state, clearly separated in energy (0.2 eV for CrI$_3$) of the first bright exciton, marked as the blue dashed line in Fig. \ref{absorbance}. In recent experiments,\cite{Seyler2018} the Stokes shift of PL emission is much lower in energy than the absorbance threshold. The reason is possibly the brightening of these dark excitons due to the electron-phonon coupling.

\begin{figure}[t]
\begin{center}
\includegraphics[width=0.7\linewidth]{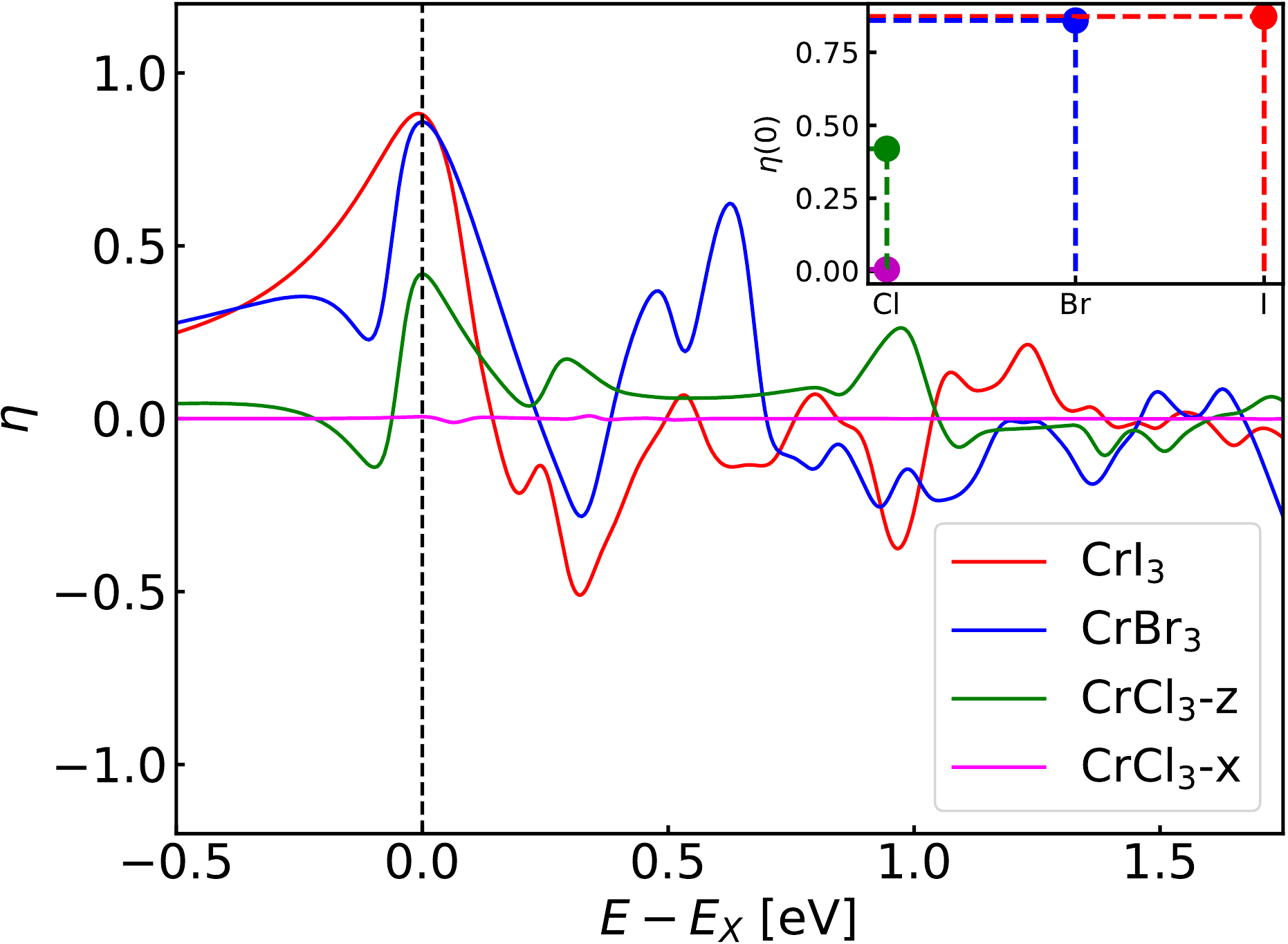}
\end{center}
\caption{Magnetic circular dichroism of the absorption, as defined in eq.
(\ref{eta}), as a function of energy, offset by the bright exciton binding
energies tabulated in \ref{gaps}, for the three compounds. For CrCl$_3$ we show
the case of both in-plane and off-plane magnetizations. The calculations include
both the GW and excitonic contributions. In the inset we show $\eta$ evaluated
at the energy of the bright exciton peak.}
\label{eta-plot}
\end{figure}

\subsection{Magnetic circular dichroism and magneto-optical  Kerr effect }
We now consider both magnetic circular dichroism in the absorption and
magneto-optical Kerr angle, which are the most common techniques to probe
the   magnetization of 2D chromium trihalides  \cite{Huang2017,Huang2018}.  In
figure \ref{eta-plot} we show the  dichroism of the absorption coefficients, defined as:
\begin{equation}
\eta\equiv \frac{A_{+}-A{_-}}{A_{+}+A{_-}},
\label{eta}
\end{equation}
for CrI$_3$, CrBr$_3$, CrCl$_3$ , all with  off-plane 
magnetization, and CrCi$_3$ with  in-plane magnetization.  The 
normalized dichroism calculated at the bright exciton energy 
shown in Fig. \ref{eta-plot} also reveals a dependence on the ligand: it is slightly larger for CrI$_3$ than CrBr$_3$, and much smaller for CrCl$_3$ with off-plane magnetization,  and almost completely negligible for in-plane magnetization.   This correlates with the spin orbit coupling of the ligand, but other factors must be at play, given the similar value obtained for CrI$_3$ and CrBr$_3$.

\begin{figure}[h!]
\begin{center}
\includegraphics[width=0.4\linewidth]{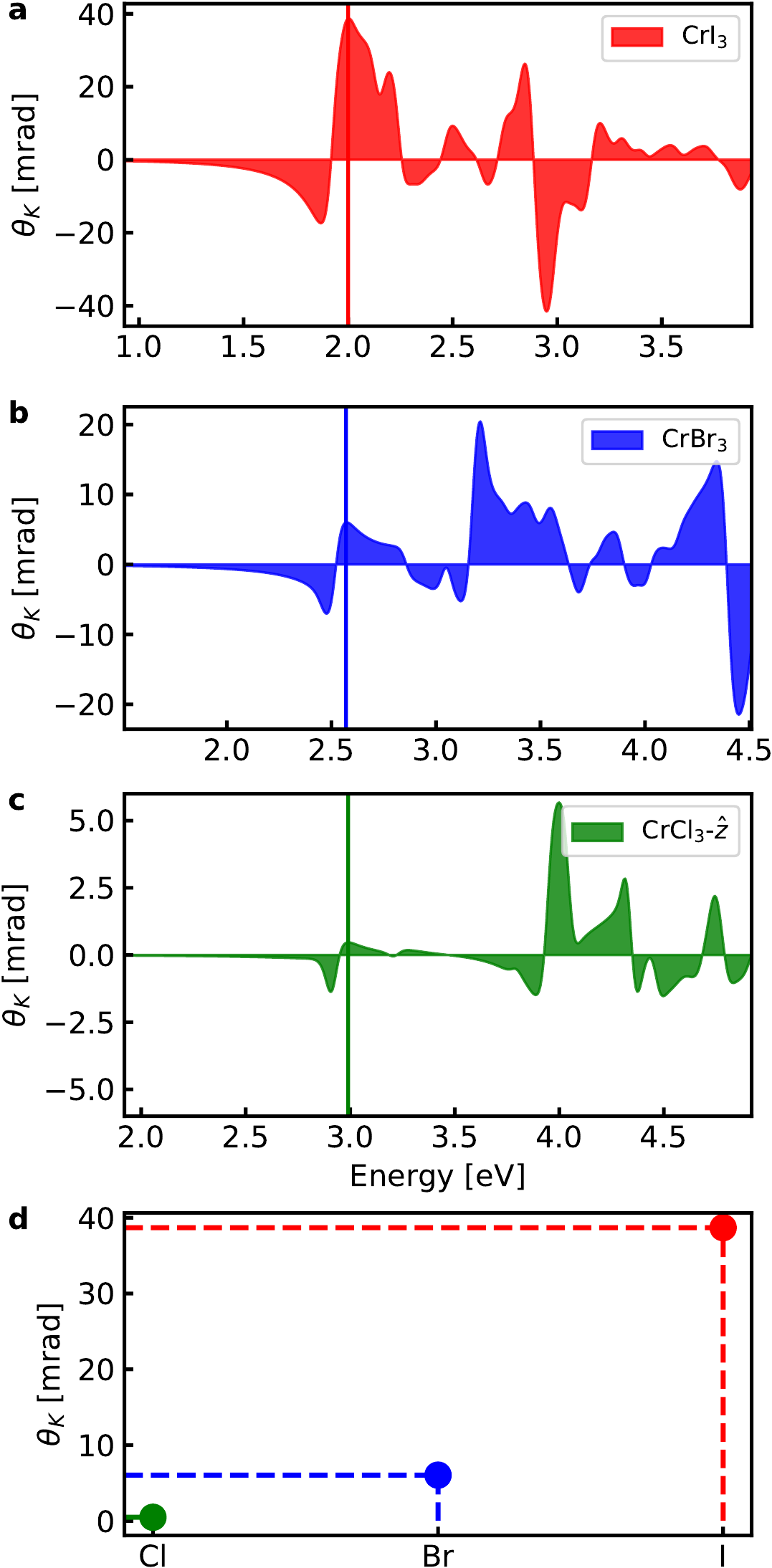}
\end{center}
\caption{Kerr spectra $\theta_K$ including excitonic effects of chromium
trihalides CrI$_3$ (red), CrBr$_3$ (blue) and CrCl$_3$ (green), all with
out-of-plane magnetization. Notice the different vertical scale. Vertical lines mark the positive threshold of Kerr spectra.}
\label{kerr-spectra}
\end{figure}

Our results for magneto-optical Kerr angle, $\theta_K$, calculated 
including both the excitonic  effects and the contribution of 
the substrate, are shown in figure \ref{kerr-spectra}(a,b,c), for the  
three CrX$_3$ compounds with off-plane magnetization. The calculation 
assumes that the 2D crystals have a finite thickness $d$, as 
established in eq. 3. Figure \ref{kerr-spectra} summarizes 
the Kerr angle spectra (in mrad). We have marked with vertical lines the positive maximum of Kerr angle in each material.
It is apparent that the Kerr angle threshold is an increasing function of the atomic weight of the ligand, as show in
figure \ref{kerr-spectra}(d). This is  expected since the magneto-optical Kerr effect arises as the interplay between magnetization, which is the same in the three compounds, and spin orbit coupling\cite{Qiu2000}, that increases for heavier ligands.

The magnitude of the Kerr angle, for the energy of the first absorption peak, computed without 
excitonic effect (see supplementary information), is very similar than the one obtained for excitons at the bright 
exciton energy. We find that this rather modest excitonic enhancement is the same for the 
three compounds.  However, as shown in Fig. \ref{absorbance}, the excitonic correction 
has a dramatic impact on the location of the absorption threshold energy

We now compare our results to experimental data. The circular polarization of
PL ($\eta$) in CrI$_3$ is 0.5 at the emission energy and the Kerr angle 8
mrad, measured at 633 nm (1.96 eV, blue-shifted 0.4 eV with respect to the
absorption threshold).\cite{Huang2017} In our calculations, our
absorption threshold is shifted with respect to experiments 0.4
eV, therefore we have to compare the experimental
results with the Kerr at 2.33 eV, obtaining 7 mrad. Nevertheless, the very
large dependence of Kerr angle on frequency at the absorption threshold would
need more experimental data for a proper comparison. In the case of CrBr$_3$ the polarization drops to
0.2.\cite{Zhang2019a} There are measurements of the magnetic anisotropy of
CrCl$_3$ monolayers and few-layers but not of the Kerr
angle.\cite{Kim2019b,Wang2019b} Our calculations capture the trend of the
polarization and the measured Kerr angle is within the range of our simulations if the
Kerr angle is measured below 2 eV.

\section{Discussion and outlook \label{disc}}

In this work we have employed state-of-the-art
\textit{ab initio} calculations including
excitonic effects to understand optical and magnetic properties of the family of
2D chromium trihalides, CrI$_3$, CrBr$_3$ and CrCl$_3$. These three materials
have the same ferromagnetic honeycomb lattice, but different ligand atoms.  We
have compared calculations  carried out both with and without excitonic effects.
We find that, for the three compounds,  the exciton binding energies  are much
larger  than the already sizeable  values found in other 2D crystals, such as
semiconducting transition metal dichalcogenides\cite{Molina-Sanchez2015}. We
find that excitons in CrI$_3$ are more extended, and thereby have a smaller binding energy than CrBr$_3$ and CrCl$_3$, reflecting a larger Cr-ligand hybridization in the iodide.

Our calculations show that the two important quantities that define the
magneto-optical response,  the magnetic circular dichroism in the absorption
$\eta$ and the Kerr angle $\theta_K$, have a marked dependence on the ligand.
This shows that, as in the case of magnetic anisotropy \cite{Lado2017}, the
ligand atom affects the strength of the magneto-optical response in Cr trihalides. Thus, the chromium atom hosts the magnetization of these materials, but the magneto-optical response is controlled by the ligand.

\section{Acknowledgements}

The computations were performed on the Tirant III cluster of the Servei
d'Inform\`{a}tica of the University of Valencia (project vlc82) and on Mare
Nostrum cluster of the Barcelona Supercomputing Center (project FI-2019-2-0034).
A. M.-S. acknowledges the Marie-Curie-COFUND program Nano TRAIN For Growth II
(Grant Agreement 713640).  G. C. acknowledges Funda\c{c}\~{a}o para a
Ci\^{e}ncia e a Tecnologia (FCT) for  Grant No. SFRH/BD/138806/2018. J. F.-R.
acknowledges financial support from  FCT for the grant
UTAP-EXPL/NTec/0046/2017, as well as Generalitat Valenciana funding Prometeo
2017/139 and MINECO-Spain (Grant No. MAT2016-78625-C2). The authors acknowledge
Claudia Cardoso and Nuno Peres for their contribution at the early stage of this work.

\bibliographystyle{apsrev-nourl}

\begin{mcitethebibliography}{66}
\providecommand*\natexlab[1]{#1}
\providecommand*\mciteSetBstSublistMode[1]{}
\providecommand*\mciteSetBstMaxWidthForm[2]{}
\providecommand*\mciteBstWouldAddEndPuncttrue
  {\def\EndOfBibitem{\unskip.}}
\providecommand*\mciteBstWouldAddEndPunctfalse
  {\let\EndOfBibitem\relax}
\providecommand*\mciteSetBstMidEndSepPunct[3]{}
\providecommand*\mciteSetBstSublistLabelBeginEnd[3]{}
\providecommand*\EndOfBibitem{}
\mciteSetBstSublistMode{f}
\mciteSetBstMaxWidthForm{subitem}{(\alph{mcitesubitemcount})}
\mciteSetBstSublistLabelBeginEnd
  {\mcitemaxwidthsubitemform\space}
  {\relax}
  {\relax}

\bibitem[Huang \latin{et~al.}(2017)Huang, Clark, Navarro-Moratalla, Klein,
  Cheng, Seyler, Zhong, Schmidgall, McGuire, Cobden, Yao, Xiao,
  Jarillo-Herrero, and Xu]{Huang2017}
Huang,~B.; Clark,~G.; Navarro-Moratalla,~E.; Klein,~D.~R.; Cheng,~R.;
  Seyler,~K.~L.; Zhong,~D.; Schmidgall,~E.; McGuire,~M.~A.; Cobden,~D.~H.;
  Yao,~W.; Xiao,~D.; Jarillo-Herrero,~P.; Xu,~X. \emph{Nature} \textbf{2017},
  \emph{546}, 270--273\relax
\mciteBstWouldAddEndPuncttrue
\mciteSetBstMidEndSepPunct{\mcitedefaultmidpunct}
{\mcitedefaultendpunct}{\mcitedefaultseppunct}\relax
\EndOfBibitem
\bibitem[Klein \latin{et~al.}(2018)Klein, MacNeill, Lado, Soriano,
  Navarro-Moratalla, Watanabe, Taniguchi, Manni, Canfield,
  Fern{\'{a}}ndez-Rossier, and Jarillo-Herrero]{Klein2018}
Klein,~D.~R.; MacNeill,~D.; Lado,~J.~L.; Soriano,~D.; Navarro-Moratalla,~E.;
  Watanabe,~K.; Taniguchi,~T.; Manni,~S.; Canfield,~P.;
  Fern{\'{a}}ndez-Rossier,~J.; Jarillo-Herrero,~P. \emph{Science (New York,
  N.Y.)} \textbf{2018}, \emph{360}, 1218--1222\relax
\mciteBstWouldAddEndPuncttrue
\mciteSetBstMidEndSepPunct{\mcitedefaultmidpunct}
{\mcitedefaultendpunct}{\mcitedefaultseppunct}\relax
\EndOfBibitem
\bibitem[Gong and Zhang(2019)Gong, and Zhang]{Gong2019}
Gong,~C.; Zhang,~X. \emph{Science (New York, N.Y.)} \textbf{2019}, \emph{363},
  4450\relax
\mciteBstWouldAddEndPuncttrue
\mciteSetBstMidEndSepPunct{\mcitedefaultmidpunct}
{\mcitedefaultendpunct}{\mcitedefaultseppunct}\relax
\EndOfBibitem
\bibitem[Gibertini \latin{et~al.}(2019)Gibertini, Koperski, Morpurgo, and
  Novoselov]{Gibertini2019}
Gibertini,~M.; Koperski,~M.; Morpurgo,~A.~F.; Novoselov,~K.~S. \emph{Nature
  Nanotechnology} \textbf{2019}, \emph{14}, 408--419\relax
\mciteBstWouldAddEndPuncttrue
\mciteSetBstMidEndSepPunct{\mcitedefaultmidpunct}
{\mcitedefaultendpunct}{\mcitedefaultseppunct}\relax
\EndOfBibitem
\bibitem[Zhang \latin{et~al.}(2019)Zhang, Shang, Jiang, Rasmita, Gao, and
  Yu]{Zhang2019a}
Zhang,~Z.; Shang,~J.; Jiang,~C.; Rasmita,~A.; Gao,~W.; Yu,~T. \emph{Nano
  Letters} \textbf{2019}, \emph{19}, 3138--3142\relax
\mciteBstWouldAddEndPuncttrue
\mciteSetBstMidEndSepPunct{\mcitedefaultmidpunct}
{\mcitedefaultendpunct}{\mcitedefaultseppunct}\relax
\EndOfBibitem
\bibitem[Kim \latin{et~al.}(2019)Kim, Kumaravadivel, Birkbeck, Kuang, Xu,
  Hopkinson, Knolle, McClarty, Berdyugin, {Ben Shalom}, Gorbachev, Haigh, Liu,
  Edgar, Novoselov, Grigorieva, and Geim]{Kim2019}
Kim,~M. \latin{et~al.}  \emph{Nature Electronics} \textbf{2019}, 1--7\relax
\mciteBstWouldAddEndPuncttrue
\mciteSetBstMidEndSepPunct{\mcitedefaultmidpunct}
{\mcitedefaultendpunct}{\mcitedefaultseppunct}\relax
\EndOfBibitem
\bibitem[Cai \latin{et~al.}(2019)Cai, Song, Wilson, Clark, He, Zhang,
  Taniguchi, Watanabe, Yao, Xiao, McGuire, Cobden, and Xu]{Cai2019}
Cai,~X.; Song,~T.; Wilson,~N.~P.; Clark,~G.; He,~M.; Zhang,~X.; Taniguchi,~T.;
  Watanabe,~K.; Yao,~W.; Xiao,~D.; McGuire,~M.~A.; Cobden,~D.~H.; Xu,~X.
  \emph{Nano Letters} \textbf{2019}, \emph{19}, 3993--3998\relax
\mciteBstWouldAddEndPuncttrue
\mciteSetBstMidEndSepPunct{\mcitedefaultmidpunct}
{\mcitedefaultendpunct}{\mcitedefaultseppunct}\relax
\EndOfBibitem
\bibitem[Gong \latin{et~al.}(2017)Gong, Li, Li, Ji, Stern, Xia, Cao, Bao, Wang,
  Wang, Qiu, Cava, Louie, Xia, and Zhang]{Gong2017}
Gong,~C.; Li,~L.; Li,~Z.; Ji,~H.; Stern,~A.; Xia,~Y.; Cao,~T.; Bao,~W.;
  Wang,~C.; Wang,~Y.; Qiu,~Z.~Q.; Cava,~R.~J.; Louie,~S.~G.; Xia,~J.; Zhang,~X.
  \emph{Nature} \textbf{2017}, \emph{546}, 265--269\relax
\mciteBstWouldAddEndPuncttrue
\mciteSetBstMidEndSepPunct{\mcitedefaultmidpunct}
{\mcitedefaultendpunct}{\mcitedefaultseppunct}\relax
\EndOfBibitem
\bibitem[Fei \latin{et~al.}(2018)Fei, Huang, Malinowski, Wang, Song, Sanchez,
  Yao, Xiao, Zhu, May, Wu, Cobden, Chu, and Xu]{Fei2018}
Fei,~Z.; Huang,~B.; Malinowski,~P.; Wang,~W.; Song,~T.; Sanchez,~J.; Yao,~W.;
  Xiao,~D.; Zhu,~X.; May,~A.~F.; Wu,~W.; Cobden,~D.~H.; Chu,~J.-H.; Xu,~X.
  \emph{Nature Materials} \textbf{2018}, \emph{17}, 778--782\relax
\mciteBstWouldAddEndPuncttrue
\mciteSetBstMidEndSepPunct{\mcitedefaultmidpunct}
{\mcitedefaultendpunct}{\mcitedefaultseppunct}\relax
\EndOfBibitem
\bibitem[Deng \latin{et~al.}(2019)Deng, Yu, Shi, Wang, Chen, and
  Zhang]{Deng2019}
Deng,~Y.; Yu,~Y.; Shi,~M.~Z.; Wang,~J.; Chen,~X.~H.; Zhang,~Y. \emph{arXiv
  preprint arXiv:1904.11468} \textbf{2019}, \relax
\mciteBstWouldAddEndPunctfalse
\mciteSetBstMidEndSepPunct{\mcitedefaultmidpunct}
{}{\mcitedefaultseppunct}\relax
\EndOfBibitem
\bibitem[Burch \latin{et~al.}(2018)Burch, Mandrus, and Park]{Burch2018}
Burch,~K.~S.; Mandrus,~D.; Park,~J.-G. \emph{Nature} \textbf{2018}, \emph{563},
  47--52\relax
\mciteBstWouldAddEndPuncttrue
\mciteSetBstMidEndSepPunct{\mcitedefaultmidpunct}
{\mcitedefaultendpunct}{\mcitedefaultseppunct}\relax
\EndOfBibitem
\bibitem[Cardoso \latin{et~al.}(2018)Cardoso, Soriano,
  Garc{\'{i}}a-Mart{\'{i}}nez, and Fern{\'{a}}ndez-Rossier]{Cardoso2018}
Cardoso,~C.; Soriano,~D.; Garc{\'{i}}a-Mart{\'{i}}nez,~N.;
  Fern{\'{a}}ndez-Rossier,~J. \emph{Physical Review Letters} \textbf{2018},
  \emph{121}, 067701\relax
\mciteBstWouldAddEndPuncttrue
\mciteSetBstMidEndSepPunct{\mcitedefaultmidpunct}
{\mcitedefaultendpunct}{\mcitedefaultseppunct}\relax
\EndOfBibitem
\bibitem[Zollner \latin{et~al.}(2019)Zollner, {Faria Junior}, and
  Fabian]{Zollner2019}
Zollner,~K.; {Faria Junior},~P.~E.; Fabian,~J. \emph{Physical Review B}
  \textbf{2019}, \emph{100}, 085128\relax
\mciteBstWouldAddEndPuncttrue
\mciteSetBstMidEndSepPunct{\mcitedefaultmidpunct}
{\mcitedefaultendpunct}{\mcitedefaultseppunct}\relax
\EndOfBibitem
\bibitem[Li \latin{et~al.}(2018)Li, Li, Du, Wang, Gu, Zhang, He, Duan, and
  Xu]{Li2018}
Li,~J.; Li,~Y.; Du,~S.; Wang,~Z.; Gu,~B.-L.; Zhang,~S.-C.; He,~K.; Duan,~W.;
  Xu,~Y. \emph{Science Advances} \textbf{2018}, \emph{5}, 5685\relax
\mciteBstWouldAddEndPuncttrue
\mciteSetBstMidEndSepPunct{\mcitedefaultmidpunct}
{\mcitedefaultendpunct}{\mcitedefaultseppunct}\relax
\EndOfBibitem
\bibitem[Sivadas \latin{et~al.}(2018)Sivadas, Okamoto, Xu, Fennie, and
  Xiao]{Sivadas2018a}
Sivadas,~N.; Okamoto,~S.; Xu,~X.; Fennie,~C.~J.; Xiao,~D. \emph{Nano Letters}
  \textbf{2018}, \emph{18}, 7658--7664\relax
\mciteBstWouldAddEndPuncttrue
\mciteSetBstMidEndSepPunct{\mcitedefaultmidpunct}
{\mcitedefaultendpunct}{\mcitedefaultseppunct}\relax
\EndOfBibitem
\bibitem[Huang \latin{et~al.}(2018)Huang, Clark, Klein, MacNeill,
  Navarro-Moratalla, Seyler, Wilson, McGuire, Cobden, Xiao, Yao,
  Jarillo-Herrero, and Xu]{Huang2018}
Huang,~B.; Clark,~G.; Klein,~D.~R.; MacNeill,~D.; Navarro-Moratalla,~E.;
  Seyler,~K.~L.; Wilson,~N.; McGuire,~M.~A.; Cobden,~D.~H.; Xiao,~D.; Yao,~W.;
  Jarillo-Herrero,~P.; Xu,~X. \emph{Nature Nanotechnology} \textbf{2018},
  \emph{13}, 544--548\relax
\mciteBstWouldAddEndPuncttrue
\mciteSetBstMidEndSepPunct{\mcitedefaultmidpunct}
{\mcitedefaultendpunct}{\mcitedefaultseppunct}\relax
\EndOfBibitem
\bibitem[Mak \latin{et~al.}(2019)Mak, Shan, and Ralph]{Mak2019}
Mak,~K.~F.; Shan,~J.; Ralph,~D.~C. \emph{Nature Reviews Physics} \textbf{2019},
  \emph{1}, 646--661\relax
\mciteBstWouldAddEndPuncttrue
\mciteSetBstMidEndSepPunct{\mcitedefaultmidpunct}
{\mcitedefaultendpunct}{\mcitedefaultseppunct}\relax
\EndOfBibitem
\bibitem[Thiel \latin{et~al.}(2019)Thiel, Wang, Tschudin, Rohner,
  Guti{\'{e}}rrez-Lezama, Ubrig, Gibertini, Giannini, Morpurgo, and
  Maletinsky]{Thiel2019}
Thiel,~L.; Wang,~Z.; Tschudin,~M.~A.; Rohner,~D.; Guti{\'{e}}rrez-Lezama,~I.;
  Ubrig,~N.; Gibertini,~M.; Giannini,~E.; Morpurgo,~A.~F.; Maletinsky,~P.
  \emph{Science} \textbf{2019}, \emph{364}, 973--976\relax
\mciteBstWouldAddEndPuncttrue
\mciteSetBstMidEndSepPunct{\mcitedefaultmidpunct}
{\mcitedefaultendpunct}{\mcitedefaultseppunct}\relax
\EndOfBibitem
\bibitem[Sun \latin{et~al.}(2019)Sun, Yi, Song, Clark, Huang, Shan, Wu, Huang,
  Gao, Chen, McGuire, Cao, Xiao, Liu, Yao, Xu, and Wu]{Sun2019}
Sun,~Z. \latin{et~al.}  \emph{Nature} \textbf{2019}, \emph{572}, 497--501\relax
\mciteBstWouldAddEndPuncttrue
\mciteSetBstMidEndSepPunct{\mcitedefaultmidpunct}
{\mcitedefaultendpunct}{\mcitedefaultseppunct}\relax
\EndOfBibitem
\bibitem[Mak \latin{et~al.}(2018)Mak, Xiao, and Shan]{Mak2018}
Mak,~K.~F.; Xiao,~D.; Shan,~J. \emph{Nature Photonics} \textbf{2018},
  \emph{12}, 451--460\relax
\mciteBstWouldAddEndPuncttrue
\mciteSetBstMidEndSepPunct{\mcitedefaultmidpunct}
{\mcitedefaultendpunct}{\mcitedefaultseppunct}\relax
\EndOfBibitem
\bibitem[Seyler \latin{et~al.}(2018)Seyler, Zhong, Huang, Linpeng, Wilson,
  Taniguchi, Watanabe, Yao, Xiao, McGuire, Fu, and Xu]{Seyler2018b}
Seyler,~K.~L.; Zhong,~D.; Huang,~B.; Linpeng,~X.; Wilson,~N.~P.; Taniguchi,~T.;
  Watanabe,~K.; Yao,~W.; Xiao,~D.; McGuire,~M.~A.; Fu,~K.-M.~C.; Xu,~X.
  \emph{Nano Letters} \textbf{2018}, \emph{18}, 3823--3828\relax
\mciteBstWouldAddEndPuncttrue
\mciteSetBstMidEndSepPunct{\mcitedefaultmidpunct}
{\mcitedefaultendpunct}{\mcitedefaultseppunct}\relax
\EndOfBibitem
\bibitem[Buchner \latin{et~al.}(2018)Buchner, H{\"{o}}fler, Henne, Ney, and
  Ney]{Buchner2018}
Buchner,~M.; H{\"{o}}fler,~K.; Henne,~B.; Ney,~V.; Ney,~A. \emph{Journal of
  Applied Physics} \textbf{2018}, \emph{124}, 161101\relax
\mciteBstWouldAddEndPuncttrue
\mciteSetBstMidEndSepPunct{\mcitedefaultmidpunct}
{\mcitedefaultendpunct}{\mcitedefaultseppunct}\relax
\EndOfBibitem
\bibitem[Argyres(1955)]{Argyres1955}
Argyres,~P.~N. \emph{Phys. Rev.} \textbf{1955}, \emph{97}, 334--345\relax
\mciteBstWouldAddEndPuncttrue
\mciteSetBstMidEndSepPunct{\mcitedefaultmidpunct}
{\mcitedefaultendpunct}{\mcitedefaultseppunct}\relax
\EndOfBibitem
\bibitem[Guo and Ebert(1995)Guo, and Ebert]{Guo1995}
Guo,~G.~Y.; Ebert,~H. \emph{Phys. Rev. B} \textbf{1995}, \emph{51},
  12633--12643\relax
\mciteBstWouldAddEndPuncttrue
\mciteSetBstMidEndSepPunct{\mcitedefaultmidpunct}
{\mcitedefaultendpunct}{\mcitedefaultseppunct}\relax
\EndOfBibitem
\bibitem[{Kumar Gudelli} and Guo(2019){Kumar Gudelli}, and Guo]{Gudelli2019}
{Kumar Gudelli},~V.; Guo,~G.-Y. \emph{New Journal of Physics} \textbf{2019},
  \emph{21}, 053012\relax
\mciteBstWouldAddEndPuncttrue
\mciteSetBstMidEndSepPunct{\mcitedefaultmidpunct}
{\mcitedefaultendpunct}{\mcitedefaultseppunct}\relax
\EndOfBibitem
\bibitem[Catarina \latin{et~al.}(2019)Catarina, Peres, and
  Fern{\'{a}}ndez-Rossier]{Catarina2019a}
Catarina,~G.; Peres,~N. M.~R.; Fern{\'{a}}ndez-Rossier, \textbf{2019}, arXiv
  preprint arXiv:1910.13371\relax
\mciteBstWouldAddEndPuncttrue
\mciteSetBstMidEndSepPunct{\mcitedefaultmidpunct}
{\mcitedefaultendpunct}{\mcitedefaultseppunct}\relax
\EndOfBibitem
\bibitem[Lado and Fern{\'{a}}ndez-Rossier(2017)Lado, and
  Fern{\'{a}}ndez-Rossier]{Lado2017}
Lado,~J.~L.; Fern{\'{a}}ndez-Rossier,~J. \emph{2D Materials} \textbf{2017},
  \emph{4}, 035002\relax
\mciteBstWouldAddEndPuncttrue
\mciteSetBstMidEndSepPunct{\mcitedefaultmidpunct}
{\mcitedefaultendpunct}{\mcitedefaultseppunct}\relax
\EndOfBibitem
\bibitem[Wu \latin{et~al.}(2019)Wu, Li, Cao, and Louie]{Wu2019a}
Wu,~M.; Li,~Z.; Cao,~T.; Louie,~S.~G. \emph{Nature Communications}
  \textbf{2019}, \emph{10}, 2371\relax
\mciteBstWouldAddEndPuncttrue
\mciteSetBstMidEndSepPunct{\mcitedefaultmidpunct}
{\mcitedefaultendpunct}{\mcitedefaultseppunct}\relax
\EndOfBibitem
\bibitem[Torelli and Olsen(2018)Torelli, and Olsen]{Torelli2018}
Torelli,~D.; Olsen,~T. \emph{2D Materials} \textbf{2018}, \emph{6}, 15028\relax
\mciteBstWouldAddEndPuncttrue
\mciteSetBstMidEndSepPunct{\mcitedefaultmidpunct}
{\mcitedefaultendpunct}{\mcitedefaultseppunct}\relax
\EndOfBibitem
\bibitem[Torelli \latin{et~al.}(2019)Torelli, Thygesen, and Olsen]{Torelli2019}
Torelli,~D.; Thygesen,~K.~S.; Olsen,~T. \emph{2D Materials} \textbf{2019},
  \emph{6}, 45018\relax
\mciteBstWouldAddEndPuncttrue
\mciteSetBstMidEndSepPunct{\mcitedefaultmidpunct}
{\mcitedefaultendpunct}{\mcitedefaultseppunct}\relax
\EndOfBibitem
\bibitem[Pervishko \latin{et~al.}(2019)Pervishko, Yudin, Gudelli, Delin,
  Eriksson, and Guo]{Pervishko2019}
Pervishko,~A.~A.; Yudin,~D.; Gudelli,~V.~K.; Delin,~A.; Eriksson,~O.;
  Guo,~G.-Y. \textbf{2019}, arXiv preprint arXiv:1909.13841\relax
\mciteBstWouldAddEndPuncttrue
\mciteSetBstMidEndSepPunct{\mcitedefaultmidpunct}
{\mcitedefaultendpunct}{\mcitedefaultseppunct}\relax
\EndOfBibitem
\bibitem[Jang \latin{et~al.}(2019)Jang, Jeong, Yoon, Ryee, and Han]{Woo2019}
Jang,~S.~W.; Jeong,~M.~Y.; Yoon,~H.; Ryee,~S.; Han,~M.~J. \emph{Phys. Rev.
  Materials} \textbf{2019}, \emph{3}, 031001\relax
\mciteBstWouldAddEndPuncttrue
\mciteSetBstMidEndSepPunct{\mcitedefaultmidpunct}
{\mcitedefaultendpunct}{\mcitedefaultseppunct}\relax
\EndOfBibitem
\bibitem[Giannozzi \latin{et~al.}(2009)Giannozzi, Baroni, Bonini, Calandra,
  Car, Cavazzoni, Ceresoli, Chiarotti, Cococcioni, Dabo, {Dal Corso},
  de~Gironcoli, Fabris, Fratesi, Gebauer, Gerstmann, Gougoussis, Kokalj,
  Lazzeri, Martin-Samos, Marzari, Mauri, Mazzarello, Paolini, Pasquarello,
  Paulatto, Sbraccia, Scandolo, Sclauzero, Seitsonen, Smogunov, Umari, and
  Wentzcovitch]{Giannozzi2009}
Giannozzi,~P. \latin{et~al.}  \emph{Journal of Physics: Condensed Matter}
  \textbf{2009}, \emph{21}, 395502\relax
\mciteBstWouldAddEndPuncttrue
\mciteSetBstMidEndSepPunct{\mcitedefaultmidpunct}
{\mcitedefaultendpunct}{\mcitedefaultseppunct}\relax
\EndOfBibitem
\bibitem[Baidya \latin{et~al.}(2018)Baidya, Yu, and Kim]{Baidya2018}
Baidya,~S.; Yu,~J.; Kim,~C.~H. \emph{Physical Review B} \textbf{2018},
  \emph{98}, 155148\relax
\mciteBstWouldAddEndPuncttrue
\mciteSetBstMidEndSepPunct{\mcitedefaultmidpunct}
{\mcitedefaultendpunct}{\mcitedefaultseppunct}\relax
\EndOfBibitem
\bibitem[Hamann(2013)]{Hamann2013}
Hamann,~D.~R. \emph{Physical Review B} \textbf{2013}, \emph{88}, 085117\relax
\mciteBstWouldAddEndPuncttrue
\mciteSetBstMidEndSepPunct{\mcitedefaultmidpunct}
{\mcitedefaultendpunct}{\mcitedefaultseppunct}\relax
\EndOfBibitem
\bibitem[van Setten \latin{et~al.}(2018)van Setten, Giantomassi, Bousquet,
  Verstraete, Hamann, Gonze, and Rignanese]{VanSetten2018}
van Setten,~M.; Giantomassi,~M.; Bousquet,~E.; Verstraete,~M.; Hamann,~D.;
  Gonze,~X.; Rignanese,~G.-M. \emph{Computer Physics Communications}
  \textbf{2018}, \emph{226}, 39--54\relax
\mciteBstWouldAddEndPuncttrue
\mciteSetBstMidEndSepPunct{\mcitedefaultmidpunct}
{\mcitedefaultendpunct}{\mcitedefaultseppunct}\relax
\EndOfBibitem
\bibitem[Olsen \latin{et~al.}(2016)Olsen, Latini, Rasmussen, and
  Thygesen]{Olsen2016}
Olsen,~T.; Latini,~S.; Rasmussen,~F.; Thygesen,~K.~S. \emph{Physical Review
  Letters} \textbf{2016}, \emph{116}, 056401\relax
\mciteBstWouldAddEndPuncttrue
\mciteSetBstMidEndSepPunct{\mcitedefaultmidpunct}
{\mcitedefaultendpunct}{\mcitedefaultseppunct}\relax
\EndOfBibitem
\bibitem[Marini \latin{et~al.}(2009)Marini, Hogan, Gr{\"{u}}ning, and
  Varsano]{Marini2009}
Marini,~A.; Hogan,~C.; Gr{\"{u}}ning,~M.; Varsano,~D. \emph{Computer Physics
  Communications} \textbf{2009}, \emph{180}, 1392--1403\relax
\mciteBstWouldAddEndPuncttrue
\mciteSetBstMidEndSepPunct{\mcitedefaultmidpunct}
{\mcitedefaultendpunct}{\mcitedefaultseppunct}\relax
\EndOfBibitem
\bibitem[Sangalli \latin{et~al.}(2019)Sangalli, Ferretti, Miranda, Attaccalite,
  Marri, Cannuccia, Melo, Marsili, Paleari, Marrazzo, Prandini, Bonf{\`{a}},
  Atambo, Affinito, Palummo, Molina-S{\'{a}}nchez, Hogan, Gr{\"{u}}ning,
  Varsano, and Marini]{Sangalli2019}
Sangalli,~D. \latin{et~al.}  \emph{Journal of Physics: Condensed Matter}
  \textbf{2019}, \emph{31}, 325902\relax
\mciteBstWouldAddEndPuncttrue
\mciteSetBstMidEndSepPunct{\mcitedefaultmidpunct}
{\mcitedefaultendpunct}{\mcitedefaultseppunct}\relax
\EndOfBibitem
\bibitem[Larson \latin{et~al.}(2013)Larson, Dvorak, and Wu]{Larson2013}
Larson,~P.; Dvorak,~M.; Wu,~Z. \emph{Physical Review B} \textbf{2013},
  \emph{88}, 125205\relax
\mciteBstWouldAddEndPuncttrue
\mciteSetBstMidEndSepPunct{\mcitedefaultmidpunct}
{\mcitedefaultendpunct}{\mcitedefaultseppunct}\relax
\EndOfBibitem
\bibitem[Rozzi \latin{et~al.}(2006)Rozzi, Varsano, Marini, Gross, and
  Rubio]{Rozzi2006}
Rozzi,~C.~A.; Varsano,~D.; Marini,~A.; Gross,~E. K.~U.; Rubio,~A. \emph{Phys.
  Rev. B} \textbf{2006}, \emph{73}, 205119\relax
\mciteBstWouldAddEndPuncttrue
\mciteSetBstMidEndSepPunct{\mcitedefaultmidpunct}
{\mcitedefaultendpunct}{\mcitedefaultseppunct}\relax
\EndOfBibitem
\bibitem[Bernardi \latin{et~al.}(2013)Bernardi, Palummo, and
  Grossman]{Bernardi2013}
Bernardi,~M.; Palummo,~M.; Grossman,~J.~C. \emph{Nano Letters} \textbf{2013},
  \emph{13}, 3664--3670\relax
\mciteBstWouldAddEndPuncttrue
\mciteSetBstMidEndSepPunct{\mcitedefaultmidpunct}
{\mcitedefaultendpunct}{\mcitedefaultseppunct}\relax
\EndOfBibitem
\bibitem[Thygesen(2017)]{Thygesen2017}
Thygesen,~K.~S. \emph{2D Materials} \textbf{2017}, \emph{4}, 022004\relax
\mciteBstWouldAddEndPuncttrue
\mciteSetBstMidEndSepPunct{\mcitedefaultmidpunct}
{\mcitedefaultendpunct}{\mcitedefaultseppunct}\relax
\EndOfBibitem
\bibitem[Constant \latin{et~al.}(2016)Constant, Hornett, Chang, and
  Hendry]{Constant2016}
Constant,~T.~J.; Hornett,~S.~M.; Chang,~D.~E.; Hendry,~E. \emph{Nature Physics}
  \textbf{2016}, \emph{12}, 124--127\relax
\mciteBstWouldAddEndPuncttrue
\mciteSetBstMidEndSepPunct{\mcitedefaultmidpunct}
{\mcitedefaultendpunct}{\mcitedefaultseppunct}\relax
\EndOfBibitem
\bibitem[McGuire \latin{et~al.}(2015)McGuire, Dixit, Cooper, and
  Sales]{McGuire2015}
McGuire,~M.~A.; Dixit,~H.; Cooper,~V.~R.; Sales,~B.~C. \emph{Chemistry of
  Materials} \textbf{2015}, \emph{27}, 612--620\relax
\mciteBstWouldAddEndPuncttrue
\mciteSetBstMidEndSepPunct{\mcitedefaultmidpunct}
{\mcitedefaultendpunct}{\mcitedefaultseppunct}\relax
\EndOfBibitem
\bibitem[McGuire(2017)]{McGuire2017}
McGuire,~M. \emph{Crystals} \textbf{2017}, \emph{7}, 121\relax
\mciteBstWouldAddEndPuncttrue
\mciteSetBstMidEndSepPunct{\mcitedefaultmidpunct}
{\mcitedefaultendpunct}{\mcitedefaultseppunct}\relax
\EndOfBibitem
\bibitem[Streltsov and Khomskii(2017)Streltsov, and Khomskii]{Streltsov2017}
Streltsov,~S.~V.; Khomskii,~D.~I. \emph{Physics-Uspekhi} \textbf{2017},
  \emph{60}, 1121--1146\relax
\mciteBstWouldAddEndPuncttrue
\mciteSetBstMidEndSepPunct{\mcitedefaultmidpunct}
{\mcitedefaultendpunct}{\mcitedefaultseppunct}\relax
\EndOfBibitem
\bibitem[Soriano \latin{et~al.}(2019)Soriano, Cardoso, and
  Fern{\'{a}}ndez-Rossier]{Soriano2019}
Soriano,~D.; Cardoso,~C.; Fern{\'{a}}ndez-Rossier,~J. \emph{Solid State
  Communications} \textbf{2019}, \emph{299}, 113662\relax
\mciteBstWouldAddEndPuncttrue
\mciteSetBstMidEndSepPunct{\mcitedefaultmidpunct}
{\mcitedefaultendpunct}{\mcitedefaultseppunct}\relax
\EndOfBibitem
\bibitem[Wang \latin{et~al.}(2011)Wang, Eyert, and
  Schwingenschl{\"{o}}gl]{Wang2011}
Wang,~H.; Eyert,~V.; Schwingenschl{\"{o}}gl,~U. \emph{Journal of Physics:
  Condensed Matter} \textbf{2011}, \emph{23}, 116003\relax
\mciteBstWouldAddEndPuncttrue
\mciteSetBstMidEndSepPunct{\mcitedefaultmidpunct}
{\mcitedefaultendpunct}{\mcitedefaultseppunct}\relax
\EndOfBibitem
\bibitem[Abramchuk \latin{et~al.}(2018)Abramchuk, Jaszewski, Metz, Osterhoudt,
  Wang, Burch, and Tafti]{Abramchuk2018}
Abramchuk,~M.; Jaszewski,~S.; Metz,~K.~R.; Osterhoudt,~G.~B.; Wang,~Y.;
  Burch,~K.~S.; Tafti,~F. \emph{Advanced Materials} \textbf{2018}, \emph{30},
  1801325\relax
\mciteBstWouldAddEndPuncttrue
\mciteSetBstMidEndSepPunct{\mcitedefaultmidpunct}
{\mcitedefaultendpunct}{\mcitedefaultseppunct}\relax
\EndOfBibitem
\bibitem[Zhang \latin{et~al.}(2015)Zhang, Qu, Zhu, and Lam]{Zhang2015}
Zhang,~W.-B.; Qu,~Q.; Zhu,~P.; Lam,~C.-H. \emph{J. Mater. Chem. C}
  \textbf{2015}, \emph{3}, 12457--12468\relax
\mciteBstWouldAddEndPuncttrue
\mciteSetBstMidEndSepPunct{\mcitedefaultmidpunct}
{\mcitedefaultendpunct}{\mcitedefaultseppunct}\relax
\EndOfBibitem
\bibitem[Jiang \latin{et~al.}(2018)Jiang, Li, Liao, Zhao, and Zhong]{Jiang2018}
Jiang,~P.; Li,~L.; Liao,~Z.; Zhao,~Y.~X.; Zhong,~Z. \emph{Nano Letters}
  \textbf{2018}, \emph{18}, 3844--3849\relax
\mciteBstWouldAddEndPuncttrue
\mciteSetBstMidEndSepPunct{\mcitedefaultmidpunct}
{\mcitedefaultendpunct}{\mcitedefaultseppunct}\relax
\EndOfBibitem
\bibitem[Molina-S{\'{a}}nchez \latin{et~al.}(2013)Molina-S{\'{a}}nchez,
  Sangalli, Hummer, Marini, and Wirtz]{Molina-Sanchez2013}
Molina-S{\'{a}}nchez,~A.; Sangalli,~D.; Hummer,~K.; Marini,~A.; Wirtz,~L.
  \emph{Physical Review B} \textbf{2013}, \emph{88}, 045412\relax
\mciteBstWouldAddEndPuncttrue
\mciteSetBstMidEndSepPunct{\mcitedefaultmidpunct}
{\mcitedefaultendpunct}{\mcitedefaultseppunct}\relax
\EndOfBibitem
\bibitem[Molina-S{\'{a}}nchez \latin{et~al.}(2015)Molina-S{\'{a}}nchez, Hummer,
  and Wirtz]{Molina-Sanchez2015}
Molina-S{\'{a}}nchez,~A.; Hummer,~K.; Wirtz,~L. \emph{Surface Science Reports}
  \textbf{2015}, \emph{70}, 554--586\relax
\mciteBstWouldAddEndPuncttrue
\mciteSetBstMidEndSepPunct{\mcitedefaultmidpunct}
{\mcitedefaultendpunct}{\mcitedefaultseppunct}\relax
\EndOfBibitem
\bibitem[Haastrup \latin{et~al.}(2018)Haastrup, Strange, Pandey, Deilmann,
  Schmidt, Hinsche, Gjerding, Torelli, Larsen, Riis-Jensen, Gath, Jacobsen,
  Mortensen, Olsen, and Thygesen]{Haastrup2018}
Haastrup,~S.; Strange,~M.; Pandey,~M.; Deilmann,~T.; Schmidt,~P.~S.;
  Hinsche,~N.~F.; Gjerding,~M.~N.; Torelli,~D.; Larsen,~P.~M.;
  Riis-Jensen,~A.~C.; Gath,~J.; Jacobsen,~K.~W.; Mortensen,~J.~J.; Olsen,~T.;
  Thygesen,~K.~S. \emph{2D Materials} \textbf{2018}, \emph{5}, 42002\relax
\mciteBstWouldAddEndPuncttrue
\mciteSetBstMidEndSepPunct{\mcitedefaultmidpunct}
{\mcitedefaultendpunct}{\mcitedefaultseppunct}\relax
\EndOfBibitem
\bibitem[Qiu \latin{et~al.}(2013)Qiu, da~Jornada, and Louie]{Qiu2013}
Qiu,~D.~Y.; da~Jornada,~F.~H.; Louie,~S.~G. \emph{Phys. Rev. Lett.}
  \textbf{2013}, \emph{111}, 216805\relax
\mciteBstWouldAddEndPuncttrue
\mciteSetBstMidEndSepPunct{\mcitedefaultmidpunct}
{\mcitedefaultendpunct}{\mcitedefaultseppunct}\relax
\EndOfBibitem
\bibitem[Latini \latin{et~al.}(2015)Latini, Olsen, and Thygesen]{Latini2015}
Latini,~S.; Olsen,~T.; Thygesen,~K.~S. \emph{Phys. Rev. B} \textbf{2015},
  \emph{92}, 245123\relax
\mciteBstWouldAddEndPuncttrue
\mciteSetBstMidEndSepPunct{\mcitedefaultmidpunct}
{\mcitedefaultendpunct}{\mcitedefaultseppunct}\relax
\EndOfBibitem
\bibitem[Noori \latin{et~al.}(2019)Noori, Cheng, Xuan, and Quek]{Noori2019}
Noori,~K.; Cheng,~N. L.~Q.; Xuan,~F.; Quek,~S.~Y. \emph{2D Materials}
  \textbf{2019}, \emph{6}, 35036\relax
\mciteBstWouldAddEndPuncttrue
\mciteSetBstMidEndSepPunct{\mcitedefaultmidpunct}
{\mcitedefaultendpunct}{\mcitedefaultseppunct}\relax
\EndOfBibitem
\bibitem[Raja \latin{et~al.}(2019)Raja, Waldecker, Zipfel, Cho, Brem, Ziegler,
  Kulig, Taniguchi, Watanabe, Malic, Heinz, Berkelbach, and
  Chernikov]{Raja2019}
Raja,~A.; Waldecker,~L.; Zipfel,~J.; Cho,~Y.; Brem,~S.; Ziegler,~J.~D.;
  Kulig,~M.; Taniguchi,~T.; Watanabe,~K.; Malic,~E.; Heinz,~T.~F.;
  Berkelbach,~T.~C.; Chernikov,~A. \emph{Nature Nanotechnology} \textbf{2019},
  \emph{14}, 832--837\relax
\mciteBstWouldAddEndPuncttrue
\mciteSetBstMidEndSepPunct{\mcitedefaultmidpunct}
{\mcitedefaultendpunct}{\mcitedefaultseppunct}\relax
\EndOfBibitem
\bibitem[Paleari \latin{et~al.}(2018)Paleari, Galvani, Amara, Ducastelle,
  Molina-S{\'{a}}nchez, and Wirtz]{Paleari2018}
Paleari,~F.; Galvani,~T.; Amara,~H.; Ducastelle,~F.; Molina-S{\'{a}}nchez,~A.;
  Wirtz,~L. \emph{2D Materials} \textbf{2018}, \emph{5}, 45017\relax
\mciteBstWouldAddEndPuncttrue
\mciteSetBstMidEndSepPunct{\mcitedefaultmidpunct}
{\mcitedefaultendpunct}{\mcitedefaultseppunct}\relax
\EndOfBibitem
\bibitem[Paleari \latin{et~al.}(2019)Paleari, {P. C. Miranda},
  Molina-S{\'{a}}nchez, and Wirtz]{Paleari2019}
Paleari,~F.; {P. C. Miranda},~H.; Molina-S{\'{a}}nchez,~A.; Wirtz,~L.
  \emph{Physical Review Letters} \textbf{2019}, \emph{122}, 187401\relax
\mciteBstWouldAddEndPuncttrue
\mciteSetBstMidEndSepPunct{\mcitedefaultmidpunct}
{\mcitedefaultendpunct}{\mcitedefaultseppunct}\relax
\EndOfBibitem
\bibitem[Seyler \latin{et~al.}(2018)Seyler, Zhong, Klein, Gao, Zhang, Huang,
  Navarro-Moratalla, Yang, Cobden, McGuire, Yao, Xiao, Jarillo-Herrero, and
  Xu]{Seyler2018}
Seyler,~K.~L.; Zhong,~D.; Klein,~D.~R.; Gao,~S.; Zhang,~X.; Huang,~B.;
  Navarro-Moratalla,~E.; Yang,~L.; Cobden,~D.~H.; McGuire,~M.~A.; Yao,~W.;
  Xiao,~D.; Jarillo-Herrero,~P.; Xu,~X. \emph{Nature Physics} \textbf{2018},
  \emph{14}, 277--281\relax
\mciteBstWouldAddEndPuncttrue
\mciteSetBstMidEndSepPunct{\mcitedefaultmidpunct}
{\mcitedefaultendpunct}{\mcitedefaultseppunct}\relax
\EndOfBibitem
\bibitem[Qiu and Bader(2000)Qiu, and Bader]{Qiu2000}
Qiu,~Z.~Q.; Bader,~S.~D. \emph{Review of Scientific Instruments} \textbf{2000},
  \emph{71}, 1243--1255\relax
\mciteBstWouldAddEndPuncttrue
\mciteSetBstMidEndSepPunct{\mcitedefaultmidpunct}
{\mcitedefaultendpunct}{\mcitedefaultseppunct}\relax
\EndOfBibitem
\bibitem[Kim \latin{et~al.}(2019)Kim, Yang, Li, Jiang, Jin, Tao, Nichols,
  Sfigakis, Zhong, Li, Tian, Cory, Miao, Shan, Mak, Lei, Sun, Zhao, and
  Tsen]{Kim2019b}
Kim,~H.~H. \latin{et~al.}  \emph{Proceedings of the National Academy of
  Sciences} \textbf{2019}, \emph{116}, 11131--11136\relax
\mciteBstWouldAddEndPuncttrue
\mciteSetBstMidEndSepPunct{\mcitedefaultmidpunct}
{\mcitedefaultendpunct}{\mcitedefaultseppunct}\relax
\EndOfBibitem
\bibitem[Wang \latin{et~al.}(2019)Wang, Gibertini, Dumcenco, Taniguchi,
  Watanabe, Giannini, and Morpurgo]{Wang2019b}
Wang,~Z.; Gibertini,~M.; Dumcenco,~D.; Taniguchi,~T.; Watanabe,~K.;
  Giannini,~E.; Morpurgo,~A.~F. \emph{Nature Nanotechnology} \textbf{2019},
  \relax
\mciteBstWouldAddEndPunctfalse
\mciteSetBstMidEndSepPunct{\mcitedefaultmidpunct}
{}{\mcitedefaultseppunct}\relax
\EndOfBibitem
\end{mcitethebibliography}
\providecommand{\latin}[1]{#1}
\providecommand*\mcitethebibliography{\thebibliography}
\csname @ifundefined\endcsname{endmcitethebibliography}
  {\let\endmcitethebibliography\endthebibliography}{}

\end{document}